\documentclass[aps,twocolumn,showpacs,preprintnumbers,amsmath,amssymb,a4paper,superscriptaddress]{revtex4-1}

\usepackage{color}
\usepackage{colordvi}
\usepackage{amssymb}
\usepackage{amsmath}
\usepackage{epsf}
\usepackage{mathrsfs}
\usepackage{graphicx}
\usepackage{appendix}
\usepackage{hyperref}
\usepackage{float}
\usepackage{tikz}
\usepackage[normalem]{ulem}

\usepackage{mwe}
\usepackage[FIGTOPCAP]{subfigure}

\usepackage{soul,xcolor}
\usepackage{pifont}
\setstcolor{red}
\definecolor{OliveGreen}{rgb}{0,0.5,0}

\newcommand{\pref}[1]{(\ref{#1})}
\newcommand{\eref}[1]{Eq.~\pref{#1}}
\newcommand{\fref}[1]{Fig.~\ref{#1}}
\newcommand{\redline}{\raisebox{2pt}{\tikz{\draw[-,red,dash dot,line width = 1.0pt](0,0) -- (10mm,0);}}}

\newcommand{\greendash}{\raisebox{2pt}{\tikz{\draw[-,green,dashed,line width = 1.0pt](0,0) -- (10mm,0);}}}

\newcommand{\blueline}{\raisebox{2pt}{\tikz{\draw[-,blue,dotted,line width = 1.0pt](0,0) -- (10mm,0);}}}

\def\be{\begin{equation}}
\def\ee{\end{equation}}
\def\bea{\begin{eqnarray}}
\def\eea{\end{eqnarray}}
\begin{document}

\title{Formation of local and global currents in a toroidal Bose{\textendash}Einstein condensate via an inhomogeneous artificial gauge field}

\author{S. Sahar S. Hejazi}
\affiliation{Department of Physics, Osaka City University, 3-3-138 Sugimoto, 558-8585 Osaka, Japan}
\email{hejazi@osaka-cu.ac.jp}

\author{Juan Polo}
\affiliation{Quantum Research Centre, Technology Innovation Institute, Abu Dhabi, UAE}

\author{Makoto Tsubota}
\affiliation{Department of Physics, Osaka City University, 3-3-138 Sugimoto, 558-8585 Osaka, Japan}
\affiliation{Department of Physics and Nambu Yoichiro Institute of Theoretical and Experimental Physics (NITEP)}
\affiliation{The OCU Advanced Research Institute for Natural Science and Technology (OCARINA), Osaka City University, 3-3-138 Sugimoto, 558-8585 Osaka, Japan}

\date{\today}

\begin{abstract}\label{abstract}

We study the effects of a position-dependent artificial gauge field on an atomic Bose--Einstein condensate in quasi-one-dimensional and two-dimensional ring settings. The inhomogeneous artificial gauge field can induce global and local currents in the Bose--Einstein condensate via phase gradients along the ring and vortices, respectively. We observe two different regimes in the system depending on the radial size of the ring and strength of the gauge field. For weak artificial gauge fields, the angular momentum increases, as expected, in a quantized manner; however, for stronger values of the fields, the angular momentum exhibits a linear (non-quantized) behavior. We also characterize the angular momentum for non-cylindrically symmetric traps. 

\end{abstract}


\maketitle
 
\section{Introduction}\label{Introduction}

Ultracold atomic gases provide excellent test-beds for studying quantum phenomena in a controlled manner. They are trapped in an external potential and cooled down to be at their lowest-energy state \cite{Ketterle_cooling,Anderson198,Barrett01,Nobel_Lec_Phillips}. As the temperature of the gas decreases below a critical value, a Bose--Einstein condensate (BEC) is obtained. In the mean-field approximation, a BEC trapped in an external potential is described by using a macroscopic wave-function that obeys a non-linear Schr\"{o}dinger-type equation, is known as the Gross-Pitaevskii equation (GPE), and has been extensively studied in various frameworks \cite{Book_pethick:08,QM_Bloch,Pitaevskii2016,Aftalion_Vortices_2001}.

Cold atomic gases have been used to investigate dynamical phenomena associated with superfluidity in low temperature physics. This involves from studying the formation of various vortex structures and their dynamics in a BEC \cite{Wu_vorticity_07,Saffman_vortex_95} in addition to more complicated nonlinear phenomena such as quantum turbulence \cite{Tsubota_QTurbulence_08,Kobayashi_QTurbulenc_07,Tsatsos_turbulence,White_vortices_14,Paoletti_QTurbulence_11}. The superfluid behavior of ultracold atomic gases, such as BECs, can cause to the appearance of topological defects, e.g., the creation of quantized vortices. In a co-rotating frame, superfluid vortex states correspond to a global energy minimum; however, their creation process usually requires to a break in the rotational symmetry at some point. 

For homogeneous rotations, there exists a critical rotational frequency for which vortices start to appear in a BEC \cite{Vortex_Madison}. This produces a sudden increase in the angular momentum and phase profile. Generally, there are three types of vortices in rotating cold atomic gases: (i) {\it visible} vortices, which can easily be detected via their density depression as well as phase winding; (ii) {\it hidden/invisible} vortices, which are not visible on the density distribution but only in the phase profile of the gas and induce a persistent current in the fluid and (iii) {\it ghost} vortices that are on the outskirt of the BEC, where the amplitude of $|\psi|^2$ is almost negligible and neither carry angular momentum nor energy \cite{Wenvortex}. The vortex nucleation and lattice formation in a rotating condensate has been studied theoretically \cite{Tsubota_vortex_2002,Fetter2009rotating} and observed experimentally \cite{Madison_Stationary,Vortex_Matthews,Vortex_Abo-Shaeer,Engels2002}.
Due to symmetry breaking, the angular momentum is transferred into the condensate through the excitation of surface modes, which results in the generation of vortices. The critical frequency is found to be greater than the vortex stability frequency \cite{Feder_Nucleation}. In a similar study the ground-state properties of a two-component BEC in a harmonic plus quadratic trap has been considered, and an artificial magnetic field was created by rotating the BEC \cite{CHEN20152193}. 

Various approaches have been employed to create vortices such as stirring a BEC with a laser beam \cite{Vortex_Madison,Madison_Stationary}, phase imprinting \cite{Vortex_Abo-Shaeer}, a rotating trap \cite{Hodby2001}, a rotating laser spoon \cite{Vortex_Madison}, rotating magnetic trap, rotating thermal cloud \cite{Williams_02}, phase engineering in two-species condensates \cite{Williams_99}, inhomogeneous synthetic magnetic fields \cite{Price_2016} and, recently, using vector gauge potential \cite{Mochol_mag,Rashi_fiber_evan,Ferris_wheel_Lemb}. The defects in BECs results in different vortex lattice structural geometries, e.g., linear vortex lattices \cite{Mcendoo_09} and zigzag arrangement of vortices \cite{LoGullo:11}.
However, for cases wherein the artificial gauge field is position-dependent, the vortex structures appear at the high intensity of magnetic field \cite{Mochol_mag,Sahar_2BEC}. This position dependence of the gauge field leads to various interesting features, such as the symmetry breaking of density profile of the two-coupled BECs \cite{Sahar_2BEC} and also the creation of a position-dependent structure of vortices in the condensate. 
In this work, we will explore the effects of an inhomogeneous gauge field on a BEC trapped in a finite-width ring geometry. This geometry is one of the simplest that is topologically non-trivial and therefore allows persistent currents (global currents) around the whole loop, while at the same time also permitting localised vortex structures. A toroidal potential can be prepared by using various methods such as: painting time-dependent potential \cite{Henderson_ring_2009} or optical lattices in the form of a ring trap with a tunable boundary phase twist, which is created by interference between the plane wave with Laguerre--Gaussian (LG) laser modes \cite{Ring_Amico}. It can also be created using a radio-frequency-dressed (RF-dressed) magnetic trap with an optical potential \cite{Heathcote_ring_2008}, or using rapidly scanned time-averaged dipole potentials \cite{Bell_ring_16}, using ring-shaped magnetic waveguide \cite{Gupta05}, and BEC can be stored in ring potential using magnetic field \cite{Arnold_06}.

Majority of the approximate theories are valid for higher dimensions however one-dimensional systems are particularly important in fundamental physics \cite{Ring_Amico,1D_Helm:15}. Since quantum effects are the strongest at low dimensions and peculiar phenomena emerge in $1$D systems \cite{Ring_Amico}, such as spin-charge separation in Luttinger liquids \cite{Morandi_Low_D}. BECs in ring traps have been used to study fundamental questions of superfluidity as well as for more practical applications such as a quantum computing playground \cite{Eckern_ring_2002}. However, in strictly $1$D systems vortices cannot exist, and therefore, the mechanisms inducing currents \cite{dobrek1999optical,polo2019oscillations} may differ from those of $2$D scenarios \cite{piazza2009vortex,mathey2016realizing}. Therefore, the transition between quasi-1D and 2D can present substantial differences when considering non-homogeneous artificial gauge fields.
Ring traps have been used in atom interferometry as a technique for realizing sensitive gravimetry \cite{Bell_ring_16}, in the application of the BEC system as a parallel of cosmological inflationary models \cite{Llorente_19} and in precision measurement devices. It has been used to study quantum fluid dynamics as well as interferometry.
The effect of a position-dependent artificial gauge field has been primarily studied in a harmonic potential \cite{Mochol_mag,Sahar_2BEC}. However, ring traps can present new phenomena owing to their symmetry and dimensionality.

This manuscript is outlines as the follows. In section II, we introduce our physical system and discuss how one can use the light--matter interactions can be used to create an artificial magnetic field in a BEC which is located near a dielectric prism. We illustrate, the results that corresponding to global in quasi-$1$D ring in section A. In section B, we illustrate the possibility of having visible and invisible vortices in a $2$D ring trap. In section III, we demonstrate how a change in the symmetry of the ring trap effects the angular momentum of the system. Finally, we conclude our observations of the effect of an inhomogeneous gauge field on a ring of BEC in section IV.

\section*{Local and global currents in ring traps} \label{sec:introduction}

Vortices (in the bulk) and persistent currents (global currents) are two hallmarks of superfluidity, and they have recently been studied extensively in atomic BECs. While vortices are fundamental objects whose the singularities of which are within the condensate field, persistent currents are usually defined in topologically non-trivial ring settings, where they appears as phase gradient along the azimuthal direction with its corresponding singularity located at the center of the trap where no condensate density exists.

In a simply connected condensate, the fundamental excitation is a vortex, which is defined by the appearance of a density minimum owing to a phase singularity in the condensate. In contrast, in a topologically non-trivial geometry, such as a ring trap, the fundamental excitation is a persistent current, which is a simple current along the azimuthal that connects onto itself after $2\pi$.

Inducing angular momentum into a cylindrically symmetric system for instance, by rotating a harmonically trapped BEC results in a collection of vortices with a winding number of unity, which are distributed through out the BEC and arrange themselves in a lattice structure that is similar to the Abrikosov lattice \cite{Abrikosov1957,Bradley2008,ORiordan2016}. This singularity in the density and the corresponding phase winding around it leads to a local current around the vortex. In contrast, considering a topologically non-trivial systems, i.e., a one-dimensional ring potential, introducing the rotation leads to a persistent current along the ring, with no visible vortices. Therefore, there is a difference in behavior of a $2$D and $1$D regime. This transition can be realized, for instance from harmonic trap to a $2$D ring and finally to a $1$D ring by creating a maximum in the center of a harmonic trap and increasing its intensity, which creates a central hole \cite{Bland2020}. There are a few methods of creating a ring trap, such as the combination of magnetic, optical, and radio-frequency fields \cite{Herve_ring:21}, or by using time-averaged adiabatic potentials (TAAPs) \cite{Navez:16}, and the use of a magnetic trap for RF-dressed atoms \cite{Heathcote:08}. 

\begin{figure}[tb]
   \centering
   \includegraphics[width=1\linewidth]{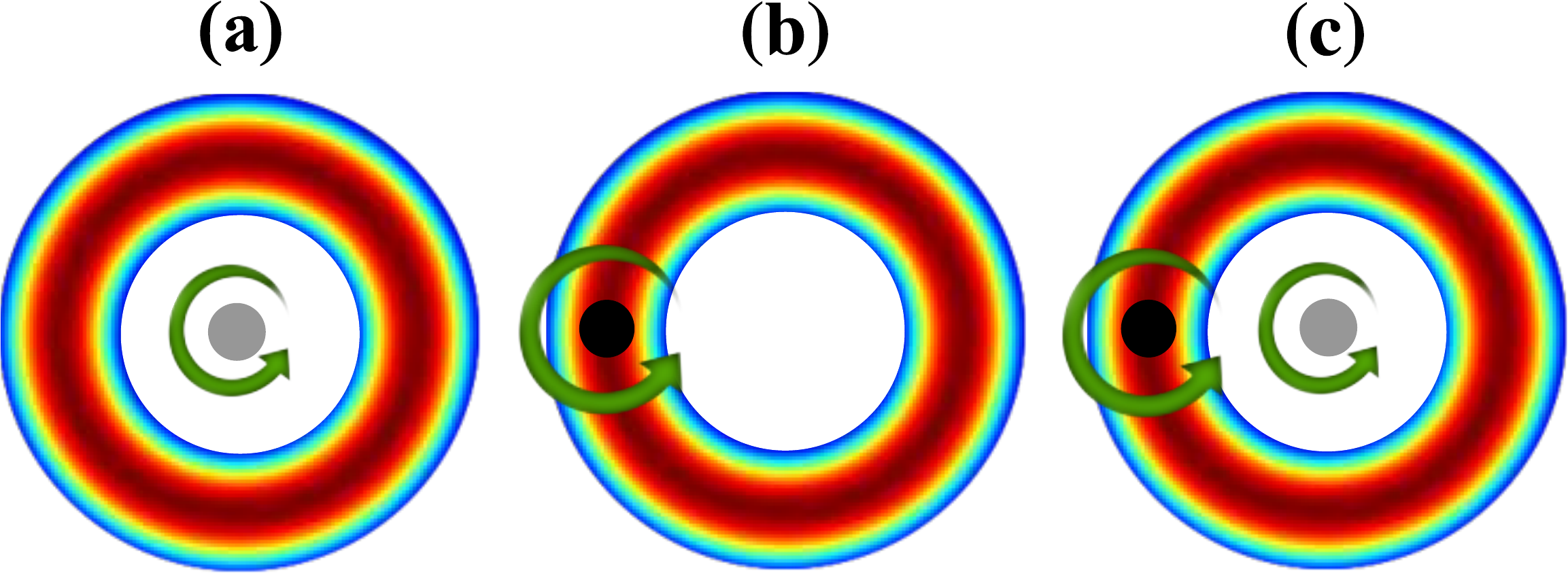}
   \caption{\textbf{Vortex configurations} Schematic of vortices in a BEC trapped in a ring potential. We consider three possible scenarios: (a) an invisible vortex (gray dot) is located at the center of the ring, and therefore the BEC experiences a global current; in plot (b) a visible vortex (black dot) results in a local current at the vicinity of where the vortex is formed; and (a) a mixture of both visible and invisible vortices. The green arrows represent an intuitive representation of the phase winding number.} \label{fig:Vortex}
\end{figure}

It can be naively expected that applying a localized (inhomogeneous) gauge field to the system results in local excitation's of rotation around the field lines; this does occur, as shown in \cite{Sahar_2BEC}, which leads to an interesting phase separation in the immiscible regime. In this work, we demonstrate that localized magnetic fields can also induce global currents.


Two types of vortices can be formed in the ring trap that experiences an inhomogeneous artificial gauge field. The first type, herein defined as visible vortices, is located in the high-density regions of the BEC cloud, and they can be observed in the density plot as a density minima, as well as in the phase (as a phase winding around their core). The other type of vortices, also known as vortex states or current states, comprise invisible vortices which form in the central part of the trap and can just be detected via phase measurements, e.g., by measuring the particle current. (see Fig.~\ref{fig:Vortex}(a)).

As vortices have an associated angular momentum around their core, they induce rotation into the BEC cloud, and the presence of visible and invisible vortices results in three different situations: (a) if the invisible vortex is at the center of the BEC ring, it induces a global current as it acts as the BEC is rotating with an external source, Fig.~\ref{fig:Vortex} (b), when the vortex is in the high-density regions of the BEC forming a density dip, it creates a local current in the BEC, as shown in Fig.~\ref{fig:Vortex}, and (c) a mixture of two types of vortices creates a mixture of global and local currents as observed in Fig.~\ref{fig:Vortex}(c).

\section{Model and Hamiltonian}\label{sec:model}

\begin{figure}[tb]
   \includegraphics[width=\linewidth]{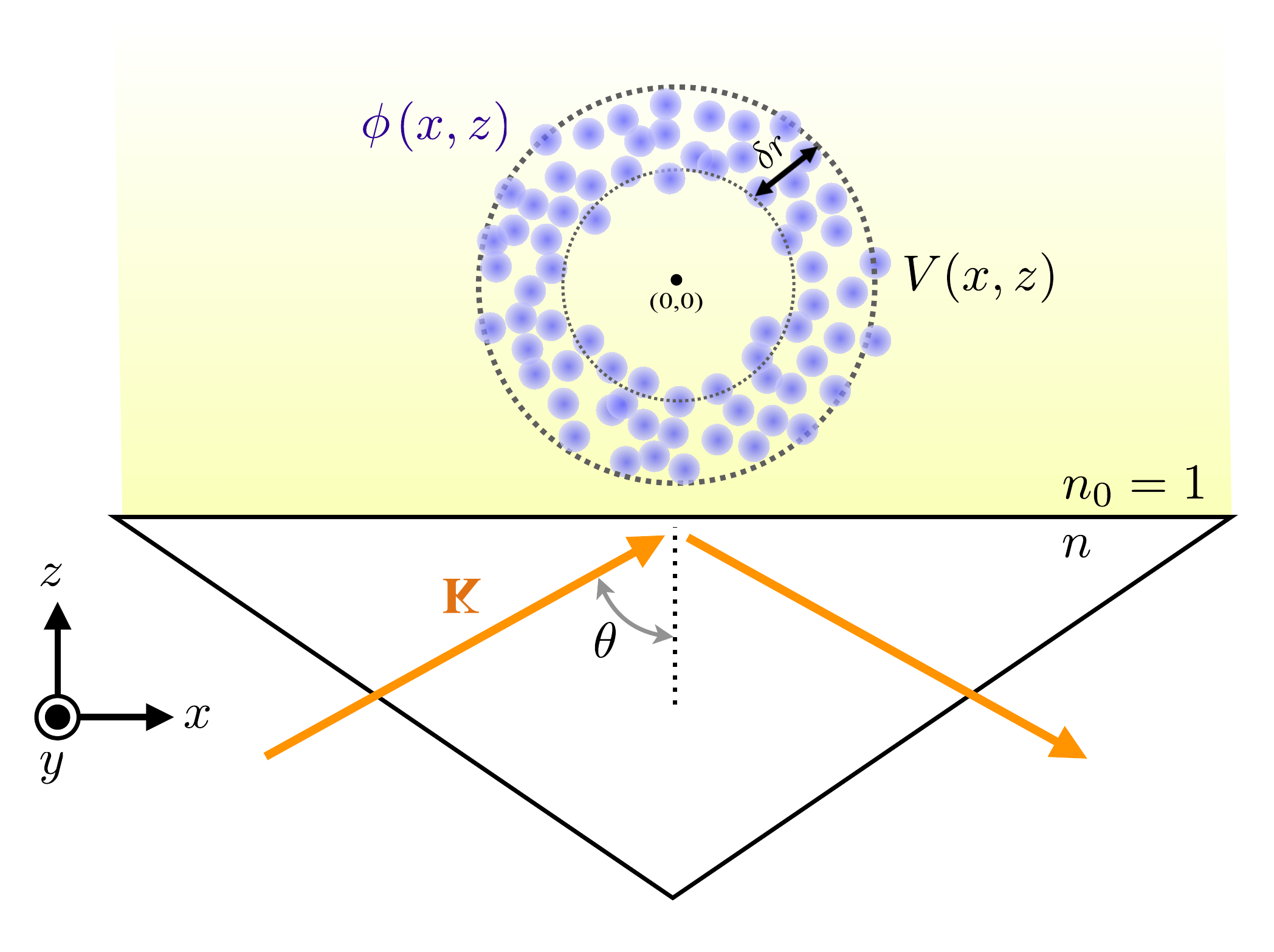}
   \caption{Schematic of a BEC cloud with the wave-function of $\phi(x,z)$ which is trapped in an external ring potential $V(x,z)$ with a thickness of $\delta r$ for the ring BEC. The cold gas is located in the vicinity of a dielectric prism having a reflective index of $n$. The orange line indicates the input laser field with wave-vector $\mathbf{k}$ and the incident angle of $\theta$.} \label{Schematic_ring}
\end{figure}

We consider a BEC of neutral alkali atoms trapped in a ring potential along the $x$--$z$ plane in the vicinity of a dielectric prism (see Fig.~\ref{Schematic_ring}). The BEC is tightly confined along the other spatial direction $y$, such that it can be effectively treated as a two-dimensional system \cite{Book_pethick:08}. The ring potential has the following form $V(x,z) = \frac{1}{2} m ( \sqrt{\omega_{x} ^2 x^{2} + \omega_{z} ^2 z^{2}} - r_0)^2$ with the frequencies of $\omega_x$ and $\omega_z$ in the $x$ and $z$ directions, respectively; $m$ is the mass of the atomic BEC, and $r_0$ corresponds to the inner radius of the ring. This ring trap is in the vacuum ($n_0=1$) just above the surface of a dielectric prism having a refractive index $n$. We choose our Cartesian coordinate system in such away that the center of the ring potential is located at the origin $(0,0)$ of the coordinate system, and the upward direction is considered as positive.

For the circular ring trap, we consider $\omega_{x} = \omega_{z}$; however in the elliptical case, we consider $\omega_{x} \neq \omega_{z}$. The BEC is created from ultra-cold alkali atoms, and each atom has two internal states: the ground state $|g\rangle$ and excited state $|e \rangle$.

\begin{figure}[tb]
   \centering
   \includegraphics[width=\linewidth]{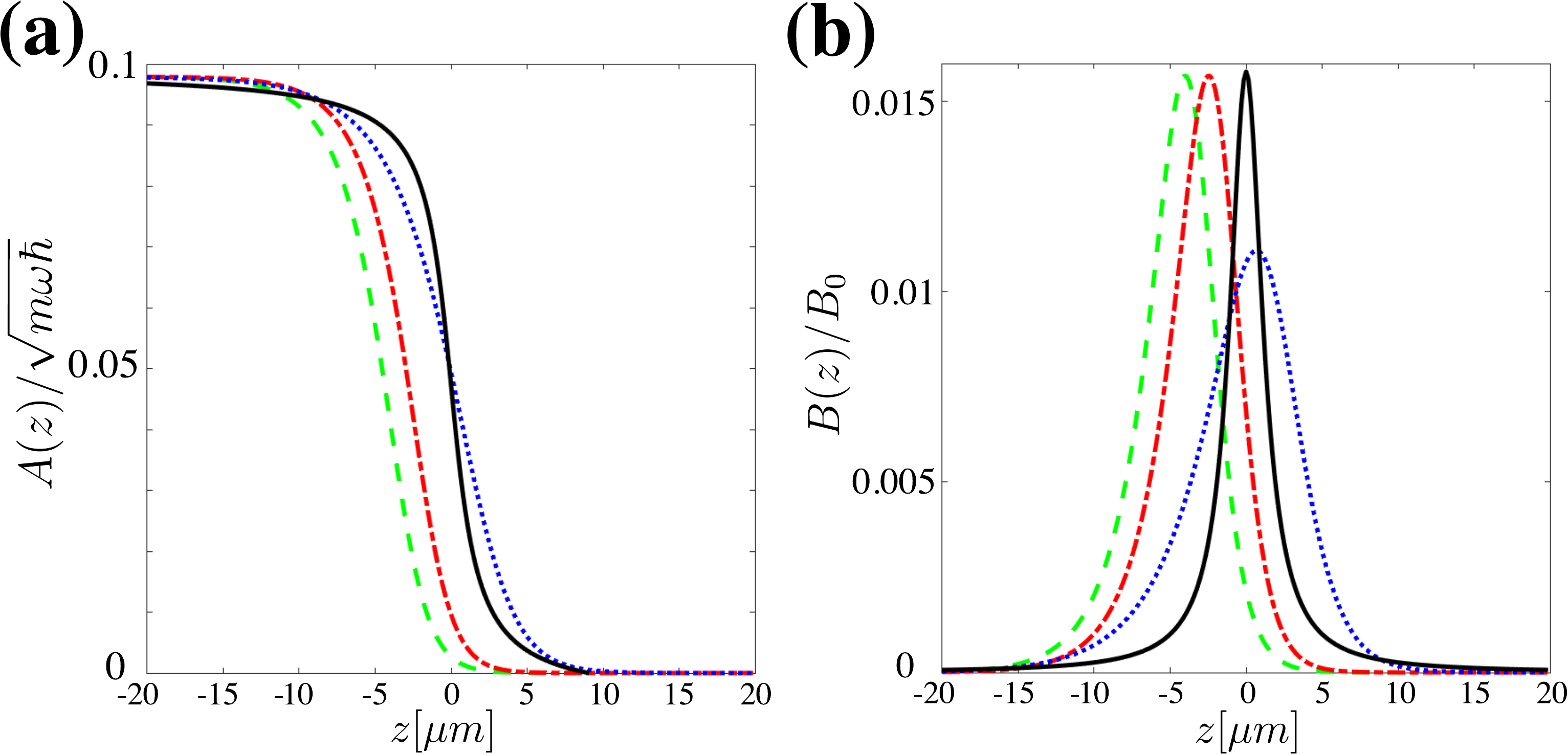}
   \caption{(a) Artificial gauge field, and (b) normalized magnetic field $B(z)/B_0$ is plotted as a function of the position $z$. The prism is located at $z=-20$. The green dashed (\protect\greendash) and red dot-dashed (\protect\redline) line correspond to an incident angle of $\theta - \theta _0 = 8 \times 10^{-4}$ rad, and $ s = 10 $ and $s=20$, respectively. The blue dotted (\protect\blueline) curve corresponds to $\theta - \theta_0 = 4 \times 10^{-4}$ rad for $ s = 20 $. The black line corresponds to our toy model 
   $A_{\text{Model}}(z) = - 0.1\left( \arctan{\left(\frac{z}{\beta}\right)} - \frac{\pi}{2} \right)/\pi$, where $\beta=1.3$. Note that $z$ is given in harmonic oscillator units.
   } \label{fig:A_B}
\end{figure}

The incident laser beam having the wave-vector $\mathbf{k}$ and frequency $\omega_{L}$, which is selected to be close to the resonance frequency of the atomic transition and propagates inside of the prism under an angle $\theta$ with respect to the interface of the prism under vacuum. The propagation angle, $\theta$, is in such a way that the beam undergoes total internal reflection (i.e., it is greater than the critical angle, $\theta_{0} = \arcsin (1/n)$), which leads to the creation of an evanescent field at the surface of the prism. The electromagnetic field, $\mathbf{E} (x,z,t)$, propagates in the $x$--$z$ plane with an amplitude $\mathbf{E}_{0}$ and decays at the surface in the positive $z$ direction with a penetration depth $d$. This can be written as $ \mathbf{E} (x,z,t) = t^{TE} (\theta) \mathbf{E}_{0} e^{- \mathrm{i} ( \omega t - \Phi (x))} e^{-z/d}$, with $t^{TE}$ as the transmission coefficient, and the running phase of $\Phi (x)$, as shown in \cite{Mochol_mag}.

The interaction between the evanescent field and the atoms in the condensate occurs via a dipole coupling, $\mathbf{d} \cdot \mathbf{E} (x,z)$, where $\mathbf{d}$ is the electric dipole moment of the atoms. For simplicity, we assume that both types of atoms have the same dipole moment and the dressed states when they interact with the light-field in the rotating wave approximation (RWA). From this dressed state, the vector potential can be calculated as follows. 
\begin{align}
	\label{A_1}
	\mathbf{A} (x,z) =  \frac{n \hbar k_{0}}{2} \left[ 1- \frac{1}{\sqrt{1+ |\frac{\kappa(x,z)}{\Delta}|^2}} \right] \sin \theta ~ \hat{\mathbf{x}},
\end{align} 
where $k_{0}$ is the amplitude of the wave-vector $\mathbf{k}$, $\Delta = \omega_{L} -\omega$ is the detuning of the laser light from the atomic resonance frequency, and the parameter $\kappa(x,z) = \mathbf{d} \cdot \mathbf{E} (x,z) / \hbar$ indicates the coupling between the electric field of the evanescent field and the dipole of atoms. The magnetic field can be calculated from the vector potential as $\mathbf{B} = \nabla \times \mathbf{A}$ \cite{Mochol_mag}.

A system that consists of a BEC cloud in the presence of the evanescent field can be described via the mean field approach with the Gross-Pitaevskii equation (GPE) as
\begin{align}
	i \frac{\partial \phi }{\partial t} = \bigg[-\frac{1}{2} ( \nabla + i \mathbf{A} )^2 + V_\text{eff} + g | \phi |^2 \bigg] \phi , \label{GPE}
\end{align}
where $\phi$ is the wave function of the BEC in the mean field description and $g$ is the coupling coefficient. In the above equation, all the parameters are scaled in harmonic oscillator units \cite{Bao_dim}. Here, we consider that the effective $V_\text{eff}$ creates a trap potential in the form of a ring.

In Fig.~\ref{fig:A_B}, we plot the artificial gauge field $A$ and magnetic field $B(z)$ versus the position $z$ for various detunings and also two different incident angles. For simplicity, we consider that the artificial gauge field of our model has the following form
\begin{equation}
   \mathbf{A}_{\text{Model}}(z) = - w_0\left( \arctan{\left(\frac{z}{\beta}\right)} - \frac{\pi}{2} \right)\hat{\mathbf{x}}, \label{eq:A}
\end{equation}
where $w_0$ changes the amplitude, and $\beta$ is the steepness of the magnetic field ($\beta \neq 0$). This results in the magnetic field $|B_{\text{Model}}(x,z)| = (\beta w_0)/(z^2 + \beta^2)$. 
The field presented in Eq.~\eqref{A_1}, reduces the number of parameters present in the experimentally realistic artificial gauge field while retaining its main features. The two parameters, $\beta$ and $w_0$, can be used to change the steepness as well as change its amplitude.

To understand how this inhomogeneous artificial gauge field affects the BEC trapped in the ring potential, we substitute the artificial gauge field $A_{\text{Model}}(z)$ from Eq.~\eqref{eq:A} into the GPE (Eq.~\eqref{GPE}) and solve the obtained equation numerically.

We compare the thickness of BEC with the healing length to define two regimes for which the quasi-one and two-dimensional ring is able to sustain vortices or not. The healing length is calculated by comparing the interaction energy ($ng$) with the kinetic energy ($n= |\psi|^2 $) and considering a fixed value for a given system \cite{HL}. We also consider $|\mathbf{A}_{\text{Model}}(x,z)|=A(x,z)$ henceforth in this paper. However, for a system that experiences an inhomogeneous artificial gauge field, the spatial scale can be written as $\xi_A = \frac{\hbar}{A(x,z) + \sqrt{2 n g}}$ (see Appendix \ref{sec:Continuity equation} for details). As the artificial the gauge field is not constant, the healing length is modified as the strength of the gauge field changes. For this calculation, we take the maximum of the $A$ field. In the limiting case where the strength of the gauge field is zero, i.e., $w_0=0.00$, the spatial scale leads to the well-known equation $\xi_0 = \hbar/ \sqrt{2ng} $, and for a strong artificial gauge field the spatial scale of the healing length is reduced. This means that vortices created by an inhomogeneous artificial gauge field will have a smaller vortex-core size as compared to homogeneous artificial gauge fields, as the typical length scale of the vortices is proportional to the local healing length of the system. Therefore, there exists competition between the size of the ring, which is fixed for a given system, and the the healing length $\xi_A$. The value of $n$ for a trapped BEC is replaced by the maximum of density as $\Bar{n}$, namely $n \approx \Bar{n}$. 

We find that vortices start to appear whenever $ \delta r \gtrsim 8\sqrt{2}\xi_A $, with $\delta r$ being the thickness of the ring, which could be analytically calculated within the Thomas--Fermi approximation, and $\sqrt{2}\xi_A$ being the typical length scale of the vortex cores in the BECs. This relation is found phenomenologically by assuming that vortices appear in pairs owing to the symmetry of the ring and field. In addition, because vortices require maxima of density surrounding them, the distance between the vortex cores as well as between the vortex and $\delta r$ should be at least $\approx 4\sqrt{2}\xi_A$ ($3\sqrt{2}\xi_A$ would actually be the minimum, but we find that a slightly larger distance is required). 

In  the following, we study the current induced by an inhomogeneous artificial magnetic that changes along $z$ in the form of an arctangent, (see \eref{eq:A}). Herein, we address how the angular momentum of this system changes as the ring transitions from one-dimensional ring, namely a ring of BEC with negligible thickness, into two-dimensional ring, i.e., a ring of BEC with substantial thickness, which has been used to study the role of dimensionality and has also been explored experimentally \cite{Eckern_ring_2002}. We also study how tilting the trapping potential and adding a small asymmetry can result in a different behavior of the angular momentum. In this work, we mainly focus on artificial gauge fields with large gradients and vary the amplitude (see \fref{fig:A_B}) as it will highlights the inhomogeneous effects of the artificial gauge field. 

\subsection{Global currents in quasi-1D ring} \label{sec:narrow_ring}

In this section, we present the results obtained from a BEC in a narrow ring potential. Here parameters of the ring trap have been selected such that the BEC in the trap can be considered as a quasi-one-dimensional BEC.

\begin{figure}[tb]
   \centering
   \includegraphics[width=1\linewidth]{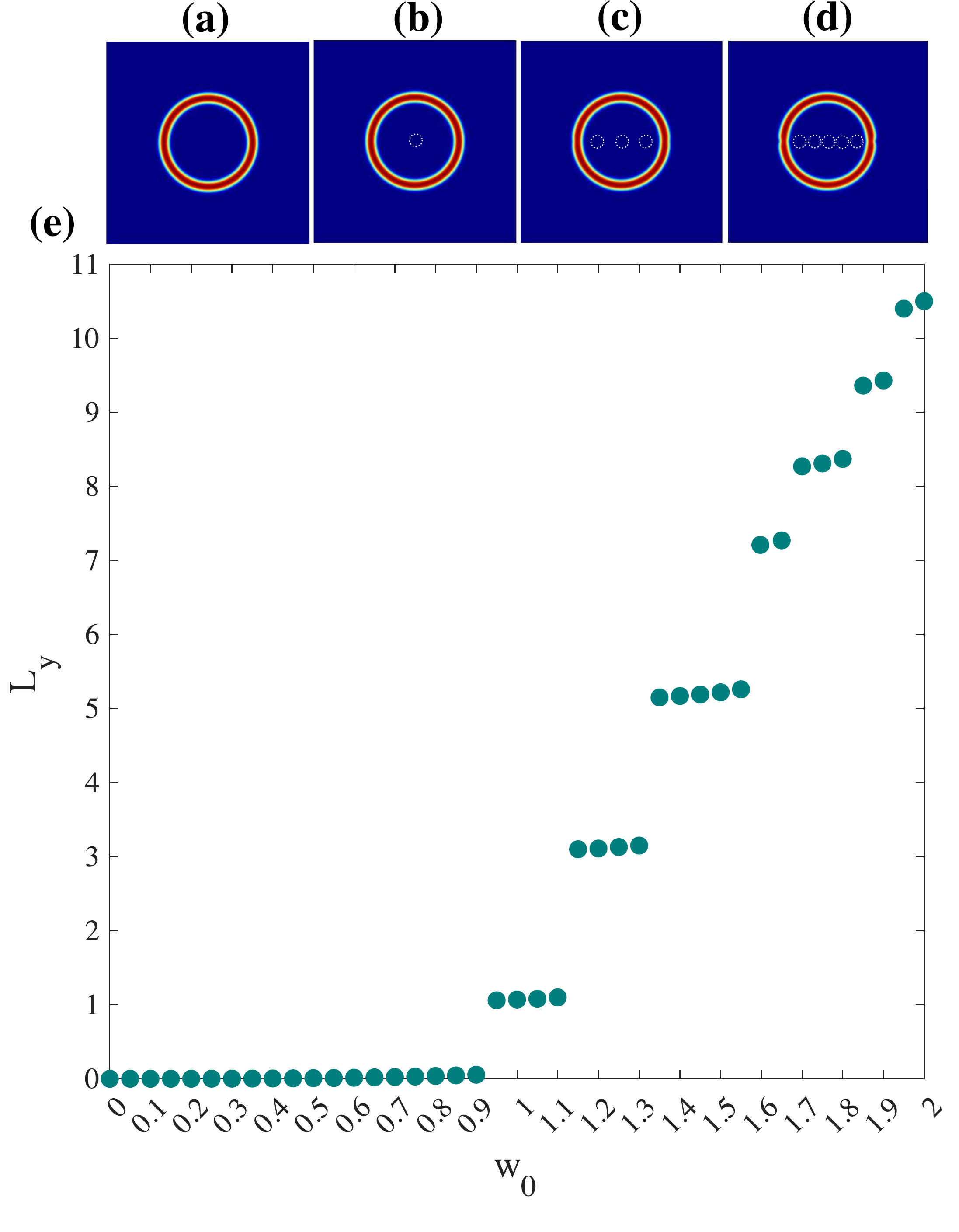} \caption{\textbf{Quantized regime for quasi-$1$D ring} The angular momentum $L_y$ of a narrow ring of the BEC is plotted with respect to the strength of the artificial gauge field $w_0$. The density plot of the BEC is shown with the location of invisible vortices for (a) $w_0=0.00$; (b) $w_0=1.00$; (c) $w_0=1.20$; (d) $w_0=1.35$ with $\beta = \frac{1}{8}$. The trapping potential has $r_0 = 6.5$ and $\eta =16$.} \label{fig:Lz_1}
\end{figure}

The number of vortices is strongly related to the strength of the artificial gauge field. In particular, by increasing the strength of the  magnetic field invisible vortices appear in the central region of the trapped BEC which induce an angular momentum in a quantized manner. This can be observed through the phase as a linearly increasing function from $0$ to multiples of $2\pi$ in the azimuthal direction. 

In Fig.~\ref{fig:Lz_1} (top row), we present the density of a ring of BEC for various values of strengths of the artificial gauge field (a) $w_0=0.00$, (b) $w_0=1.00$, (c) $w_0=1.20$, and (d) $w_0=1.35$ for the BEC with the coupling coefficient of $g=800$. We consider a ring trap which has the following potential $V_\text{eff} = \frac{1}{2} v_0 ^2 (\eta \sqrt{x^{2} + z^{2}} -r_0)^2$ 
with $r_0 = 6.5$ and $\eta =16$ and the artificial gauge field has $\beta = \frac{1}{4}$. 
Figure~\ref{fig:Lz_1} (e) displays the angular momentum, $L_{y} = (\mathbf{r} \times \mathbf{p})_y = -i\hbar (z \partial_x - x \partial_z)$, versus the strength of the artificial gauge field $w_0$. 
Panel (a), where $w_0=0.00$, does not have any artificial gauge field, therefore, presents no current in the ring. For a higher value of $w_0$ although the density plot does not change significantly, the presence of the invisible vortices results in a change in the global current. Here we show the location of the invisible vortices with doted circles.
In panel (b), where $w_0=1.00$, there exists just one invisible vortex, which results in a global current with the unit angular momentum and phase of $2\pi$. Due to the single-axis dependence of the artificial gauge field, $z$, and the symmetry of the trapping potential, as the strength of the gauge field increases, the appearance of invisible vortices do not always appear in steps of one. For example, at $w_0=1.20$ in panel (c), we see that two invisible vortices come from left and right into the system, thus to three invisible vortices which can be detected from phase plot with $6 \pi$. Finally, we note that, as discussed above, for large values of $\omega_0$ the healing length will be largely reduced and it is expected that vortices will start to appear even for thin rings (which still retain a 2D character).

\subsection{Visible and invisible vortices in quasi-2D rings}\label{sec:visible_invisible}

\begin{figure}[tb]
   \centering
   \includegraphics[width=1\linewidth]{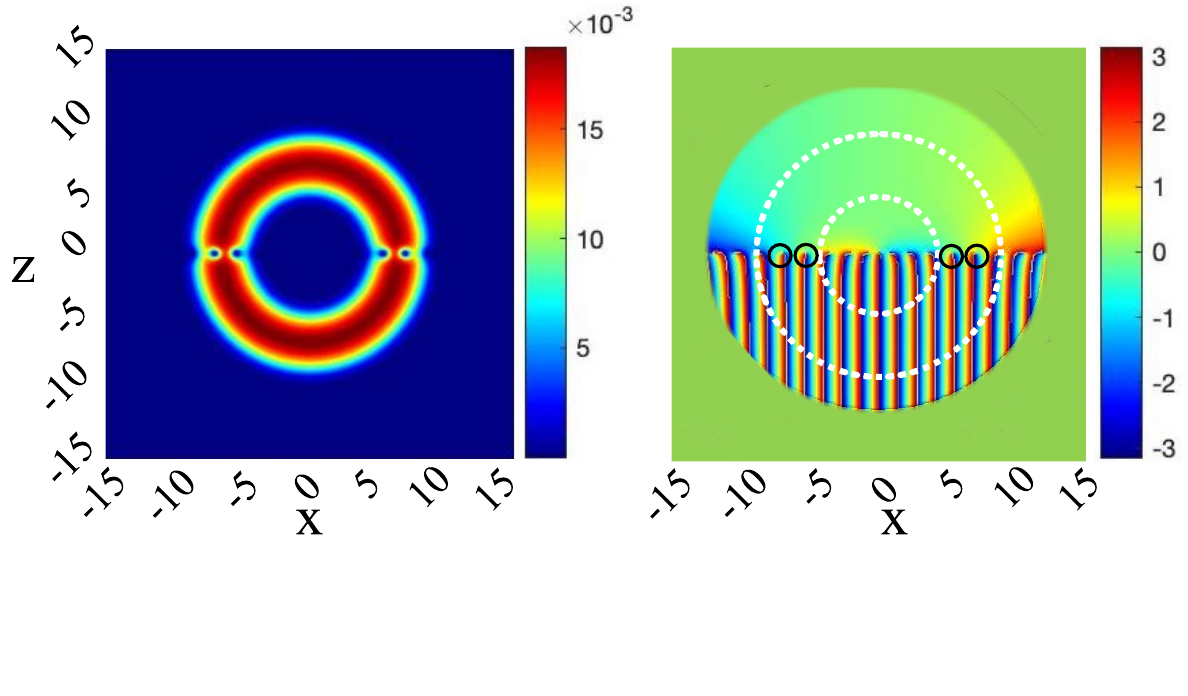}
   \caption{\textbf{Visible and invisible vortices} The density (left figure) and phase (right figure) of the BEC that trapped in a ring potential of parameters $v_0=1$, $\eta=0.9$, and radius of $r_0=6.5$. The strength of the gauge field is $w_0=1.5$ with $\beta=1/8$. The white dash in the phase plot shows the position of the trap, and the black circles indicate the locations of visible vortices.} \label{fig:2d_ring}
\end{figure}

In this section, we consider a wider potential (a ring potential with thicker radius), i.e., a trapping potential with $v_0=1$, $\eta=0.9$, and radius of $r_0=6.5$. As the thickness of the ring of the BEC, $\delta r_{\text{ring}}$, is substantial as compared to the healing length $\xi$, therefore the trap can hold visible vortices in it. As shown in Fig.~\ref{fig:2d_ring}, at a high value of the gauge field for example $w_0=1.50$, visible vortices appear in the BEC, in addition to invisible vortices in the low density. The presence of the invisible vortices here is the result of the inhomogeneous gauge field, which can be observed in the phase plot. Indeed, in Fig.~\ref{fig:2d_ring}(d), there are four invisible vortices in the ring of the BEC that can be found by calculating the gradient of the phase through $L_y$.

\begin{figure}[tb]
   \centering
   \includegraphics[width=1\linewidth]{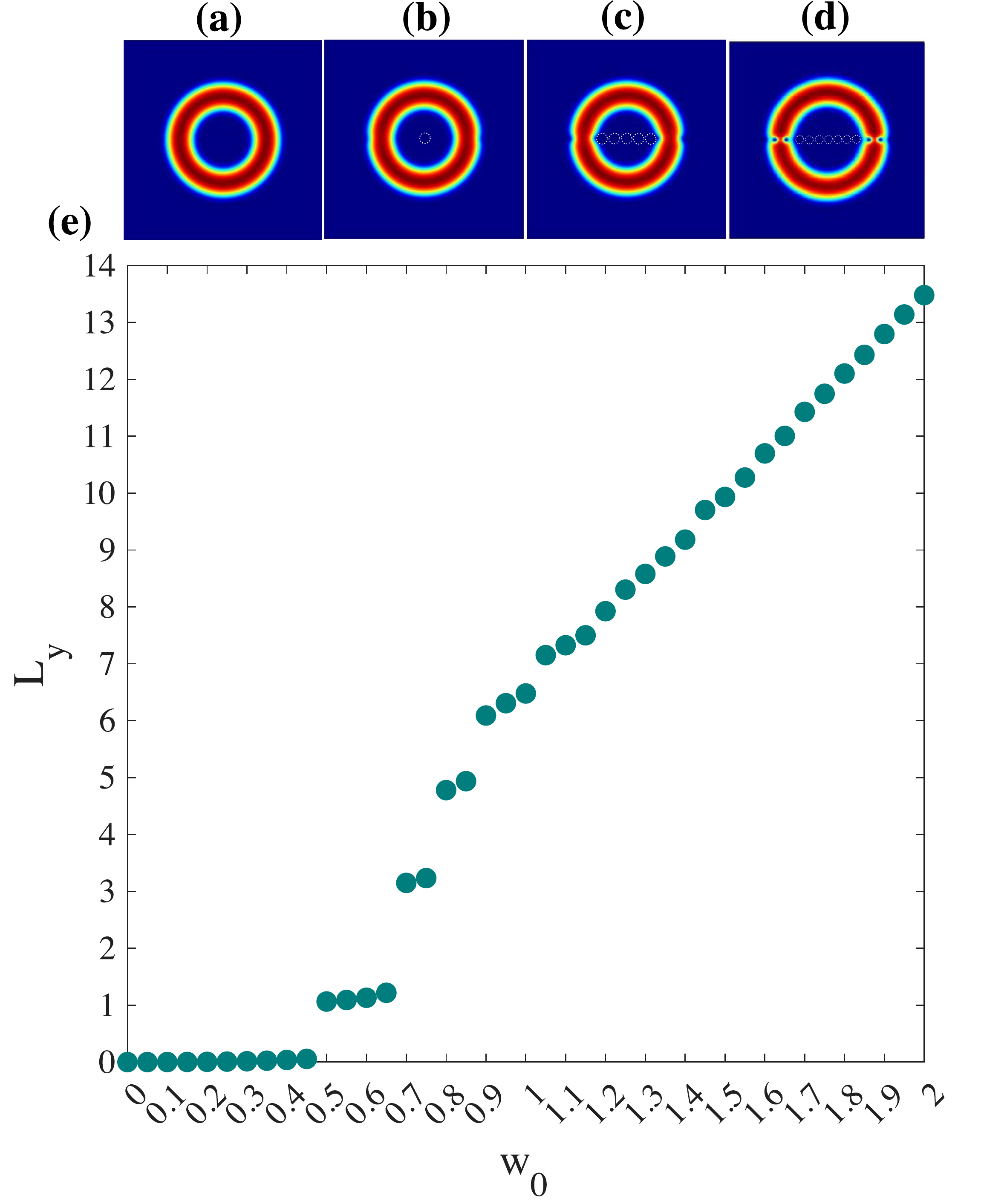}
   \caption{\textbf{Quantized and linear regimes of a $2$D ring} The density plot of the ground state of a ring of the BEC with the parameters $v_0=1$, $\eta=0.9$, and radius of $r_0=6.5$ that has a gauge field strength of (a) $w_0=0.00$; (b) $w_0=0.50$; (c) $w_0=0.80$; (d) $w_0=1.50$ with $\beta=1/8$. Panel (a) shows the density of the ground state of the cloud of the BEC when there is no artificial gauge field. As the strength of the gauge field increases, at first the invisible vortices enter the system, as can be observed in panel (b) and angular the angular momentum increases by quanta of one. In panel (c), there are five invisible vortices in the BEC which leads to a $10 \pi$ phase. Finally, in panel (d), visible vortices enter the BEC and the angular momentum is $20\pi$. Figure (e) presents the angular momentum of $L_y$ with respect to the strength of the artificial gauge field $w_0$. This plot shows both quantize and linear regimes of the angular momentum. For comparison, please refer to the angular momentum plot of the harmonic trap in Fig.~\ref{fig:HO}, as it exhibits same behavior.}  \label{fig:ch_ring_1}
\end{figure}

In this regime of the trap thickness, two different types of behavior of the angular momentum are observed, as shown in Fig.~\ref{fig:ch_ring_1}: quantized and linear regimes. When increasing the strength of the artificial gauge field, the system first presents invisible vortices. However, on further increasing the artificial gauge field, visible vortices start to appear into the high-density regions of the BEC, (see Fig.~\ref{fig:ch_ring_1}, at $w_0=0.8$). Moreover, as the the strength of the gauge field increases, the angular momentum $L_{y}$ starts to increase linearly with respect to the strength of the gauge field (see \fref{fig:ch_ring_1}). It should be noted that the same symmetry argument, discussed at the end of the previous sections, can also be made here regarding the increase in the two units of angular momentum.

This transition from a quantized angular momentum to a linear regime can also be observed in harmonically trapped BECs (see Fig.~\ref{fig:HO} in the Appendix). Therefore, it is an effect that originates in the inhomogeneity of the artificial gauge field rather than the geometry of the trapping potential. In the case of ring traps, we estimate that this linear behavior starts to appear whenever the thickness of the ring of BEC, $\delta r$ is comparable with the healing length, $\xi_A$. %
Therefore, vortices start to appear in the bulk, and the healing length is substantially affected by the artificial gauge field. In the harmonic trap, however, vortices start to appear at smaller values of $w_0$ (see Appendix A) in the high density region (center of the trap), and thus, the linearization of $L_y$ also occurs sooner. 
In our simulation, the maximum density of the ground state for the harmonic trap considered in the appendix is $\Bar{n} = 4.4 \times 10^{-3}$, in natural units ($\hbar = m = 1 $) with interaction strength of $g=800$; in the case of the ring trap, we have $\Bar{n} = 7.9 \times 10^{-3}$, and consequently, $\xi_{0,\text{ring}} \approx 0.28$.

\section{Symmetry breaking in the elliptical ring trap} \label{sec:Elliptical_ring}

We have mentioned, in Sec.~\ref{sec:visible_invisible}, that the system is highly sensitive to the symmetries of the ring and artificial gauge field. Therefore, in this section we analyze how a small asymmetry in the trapping potential can affect measurable quantities such as the angular momentum. Instead of considering a perfect circular ring, we consider a more realistic situation with a slightly deformed elliptical ring.

\begin{figure}[!]
   \centering
   \includegraphics[width=0.95
   \linewidth]{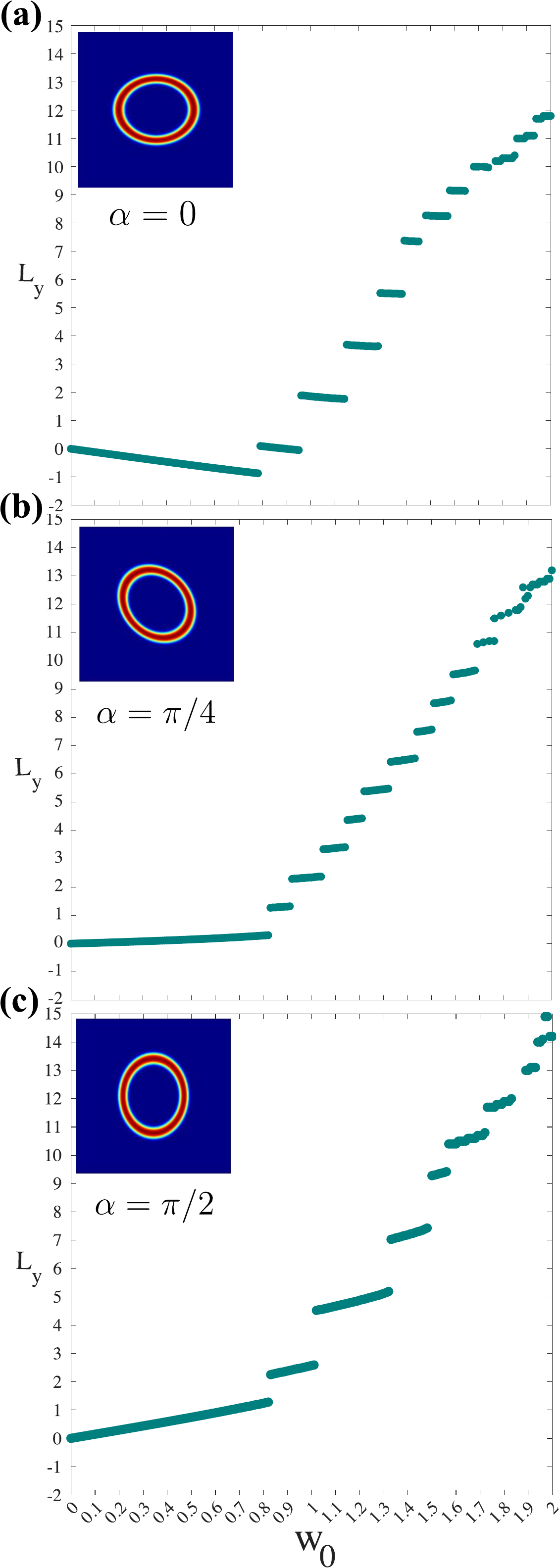}
   \caption{\textbf{Symmetry breaking of angular momentum} The angular momentum of an elliptical rings as shown in Eq.~\eqref{eq:ellipse} with $\eta=0.9$, $r_0=6.5$, $\gamma=\pi/2$, and $v_0=4$, with the ellipticity of $(1,1.5)$ is plotted for various angles: (a) $\alpha=0$, (b) $\alpha=\pi/4$, and (c) $\alpha=\pi/2$.} \label{fig:Lz_3}
\end{figure}

In particular, we look at the response of the angular momentum to different tilting angles of the major axis of an elliptical trap and compare it with the circular case. Moreover, for completeness, we also compare this with a homogeneous gauge field that represents a global rotation in the Appendix C. 

The equation of an ellipse with a semi-major axis $a$ and semi-minor axis $b$, the equation is given by $x^2/a^2 +z^2/b^2=1$, where the eccentricity of the ellipse is $e=\sqrt{1-a^2/b^2}$. By changing the eccentricity $e$ of the ring trap from a circle with $e=1$ to an elliptical with $e < 1$ trap and then tilting the major axis, one can observe that the angular momentum of the system changes as the tilting changes. In order to take into consideration the tilting of the ellipse with respect to the $x$ coordinate, we replace $x$ by $x \cos \gamma$ and $y$ by $y \sin \gamma$, and the potential thus becomes
\begin{equation}
  V(x,z) = \frac{v_0 ^2}{2} (\eta \sqrt{\frac{x^{2}}{a^2 } \cos^2 \gamma+ \frac{z^{2}}{b^2} \sin^2 \gamma} - r_0 ^2). \label{eq:ellipse} 
\end{equation}

Fig.~\ref{fig:Lz_3} shows that the angular momentum of a BEC confined in a tight elliptical trap is considerably affected by the change in the axial line of the trap with respect to line of the high-density line of the magnetic field.  In each figure, the angular momentum of the elliptical ring BEC  with the eccentricity of $e=0.745, (a = 1, b = 1.5)$ is plotted for (a) the tilting angle $\alpha=0$, (b) $\alpha = \pi /4$, and (c) $\alpha = \pi/2$ with $\eta=0.9$, $r_0=6.5$, and $\gamma=\pi/2$. 
We show that, for elliptical traps, the angular momentum does not present flat-top steps, but they have a linear dependence on the artificial magnetic field with a slope that depends on the tilting angle of the ellipse $\alpha$ (see Figure \ref{fig:Lz_1} for comparison, as the angular momentum in the quantized regime increases by an integer number).

In this figure, we observe that the angular momentum does not always increase in a quantized manner nor linearly with respect to the increase in the strength of the artificial gauge field, in which it depends on ellipticity and the angle of tilting of the trapping potential with respect to inhomogeneous artificial gauge field. 

For instance, as shown in Fig.~\ref{fig:Lz_3} (a), the angular momentum steps of the elliptical trap with a tilting angle of $\alpha=0$, in comparison with the circular flat steps, decreases linearly as the strength of the artificial gauge field increases.
However, for $\alpha = \pi/ 2$, the situation is opposite, as $w_0$ increases the plateaus of the angular moment increase linearly. All this occurs while the angular momentum presents quantized-like jumps of total angular momentum $L_y$. We also note that, for a tilting angle of $\alpha = \pi/4$, the angular moment of the elliptical ring of the BEC exhibits the same behavior as that obseved for the circular potential with almost flat plateaus.

Quite interestingly, such behavior is not observed inn the case of homogeneous artificial gauge fields such as rotations (see Appendix A). Owing to the quantized circulation of superfluids and the continuity of the wavefunction, this can be expected that the angular momentum would does not depend on the particular path being enclosed. Furthermore, for global rotations, the current appears to be independent of the tilting angle of the ellipse. Thus, we conclude that, again, there is sensitivity to the symmetries produced by the inhomogeneous gauge field in the elliptical ring trap (see discussion on the closed path integral in Appendix B). 

\section{Summary and conclusions}\label{sec:conclusion}

We showed how a position-dependent artificial gauge field has a fundamentally different impact on cold atomic gases in comparison to the case of homogeneous rotation. In our system, this difference between rotation and artificial gauge field appears as local rotations in addition to persistent currents (global rotation) simultaneously. 

We investigated the behavior of the ground-state of the BEC, which is trapped in a ring-shaped potential at the vicinity of a dielectric prism. An artificial magnetic field is produced due to the interaction between the cold atomic gases and the evanescent field that emerges at the interface between the prism and vacuum area when the laser beam goes undergoes the total internal reflection. As this system has sufficient degrees of freedom, we monitor the formation of ghost, invisible, and visible vortices by varying the strength of the gauge field. 

The position-dependent artificial magnetic field introduces a local angular momentum into this system, which results in the inhomogeneous formation of vortex structure in the BEC. This leads to creation of a vortex line in the BEC as it is aligned along the surface of the prism in vacuum. The number and locations of the vortices depends on the detuning and incident angle of input laser beam. However, originally the formation of the vortices was originally owing to the rotation of the BEC with particular frequency; however, this is due to the local injection of angular momentum owing to the artificial gauge field, to study this, we solve the GPE in two dimensions.

This open new possibilities to study superfluid systems in different artificial gauge fields, as the position-dependent artificial gauge field can produce local and global currents in the BEC. Undoubtedly, studying of the dynamics of this system would be interesting in order to compare the effect of visible and invisible vortices on the time evolution of the BEC in the ring. Another application could be for measuring the position-dependent artificial gauge field using the rotation or vortices imparted into the system. Furthermore, to study the complex nonlinear behavior of a quantum system such as the quantum turbulence, generally, the ground state of the system contains vortices distributed homogeneously across the BEC, and the dynamic of the system exhibits some nonlinear behavior of this system. However, in this proposal, due to the position-dependent of the artificial gauge field, vortices were formed with a linear structure.

Acknowledgement: We are immensely grateful to Thomas Busch for his comments on an earlier version of the manuscript and for our helpful discussions. We also would like to thank Osaka City University and Quantum Research Centre. JP and MT acknowledge the funding from the Japan Society for the Promotion of Science KAKENHI Grant No.~20K14417 and No.~20H01855, respectively.

\appendix
\section{Harmonic trap}\label{App_A}

We plot the density of the ground state of the BEC which is trapped in a harmonic potential as well as the change in the angular momentum with respect to the artificial gauge field strength. We use harmonic potential to show that the transition from the step to the linear regime that has been observed in Fig.~\ref{fig:ch_ring_1} also occurs in the harmonic potential, and consequently, it is not a result of the shape of the potential. We observe a similar pattern in change of angular momentum as $w_0$ increases, at the beginning, the angular momentum of the system increases in steps and after a certain increase in the gauge field ($w_0 \approx 0.50$), it enters the linear regime. Furthermore, the healing length is $\xi_{0} \approx 0.38$, and decreases as the strength of the gauge field increases.

\begin{figure}[tb]
   \centering
   \includegraphics[width=1\linewidth]{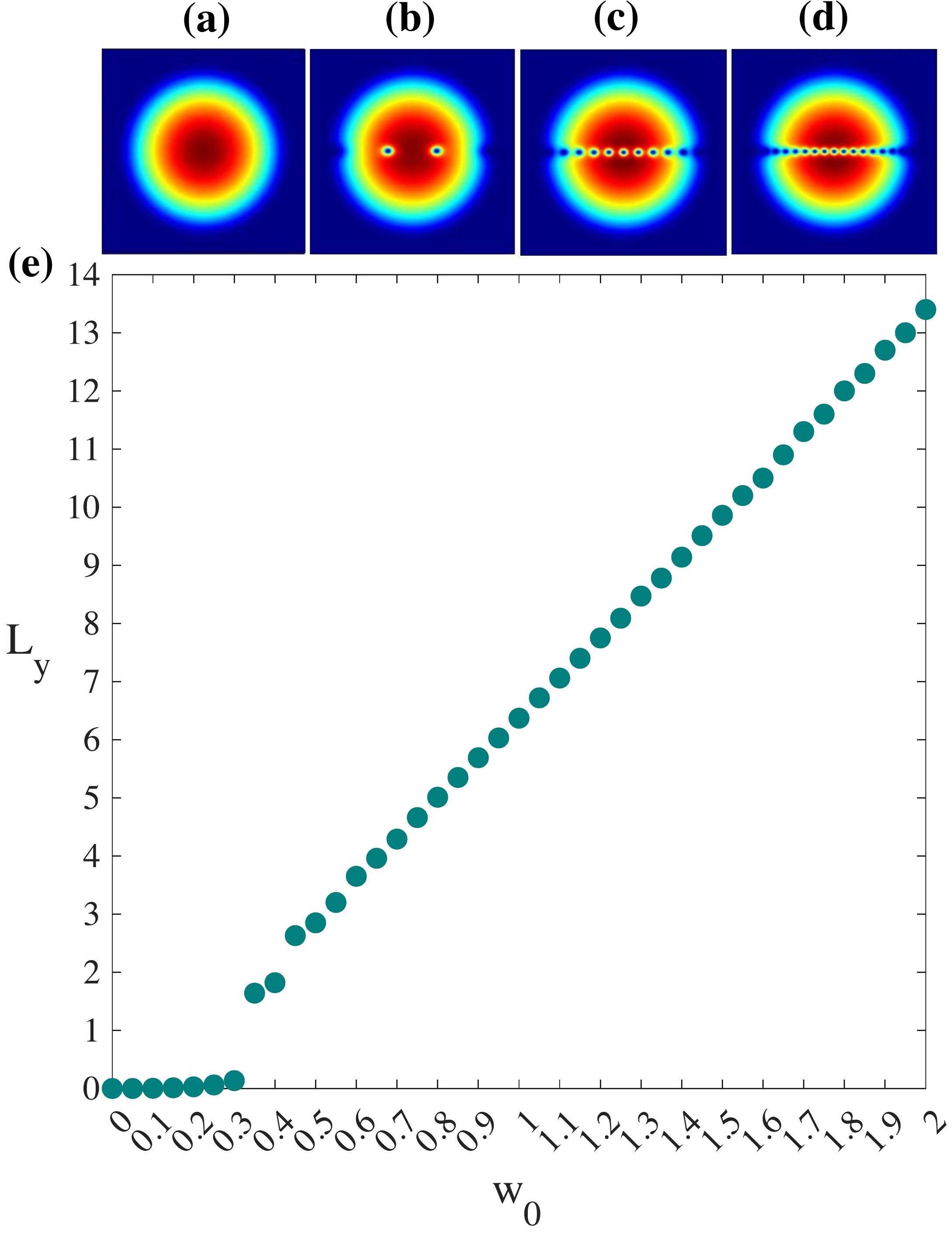}
   \caption{{\textbf{Harmonic oscillator} Panels (a)--(d) show the density plot of a BEC trapped in a harmonic potential $V= (\frac{1}{6.5})^2(x^2 + z^2);$. The artificial gauge field has the same value as Fig.~\ref{fig:ch_ring_1} with $v_0=1$ and $\eta=0.9$. The strength of the gauge field is (a) $w_0=0.00$; (b) $w_0=0.40$; (c) $w_0=1.00$; and (d) $w_0=1.50$. In plot (e) the angular momentum $L_y$ is plotted with respect to the gauge field strength $w_0$.} }\label{fig:HO}
\end{figure}

The two distinguished regimes, i.e., quantized and linear, for the angular momentum is a results of the fact that the amount of angular moment added by one vortex depends not only on the number of vortices but also its size, as the size of the vortices is reduced as the strength of the gauge field increases. In other words as the strength of angular momentum increases, the healing length reduces, consequently the amount of the angular momentum that is added by one vortex to BEC changes (see Appendix \ref{sec:Continuity equation}).

\section{Global induced rotation}\label{App_B}

\begin{figure}[tb]
   \centering
   \includegraphics[width=1\linewidth]{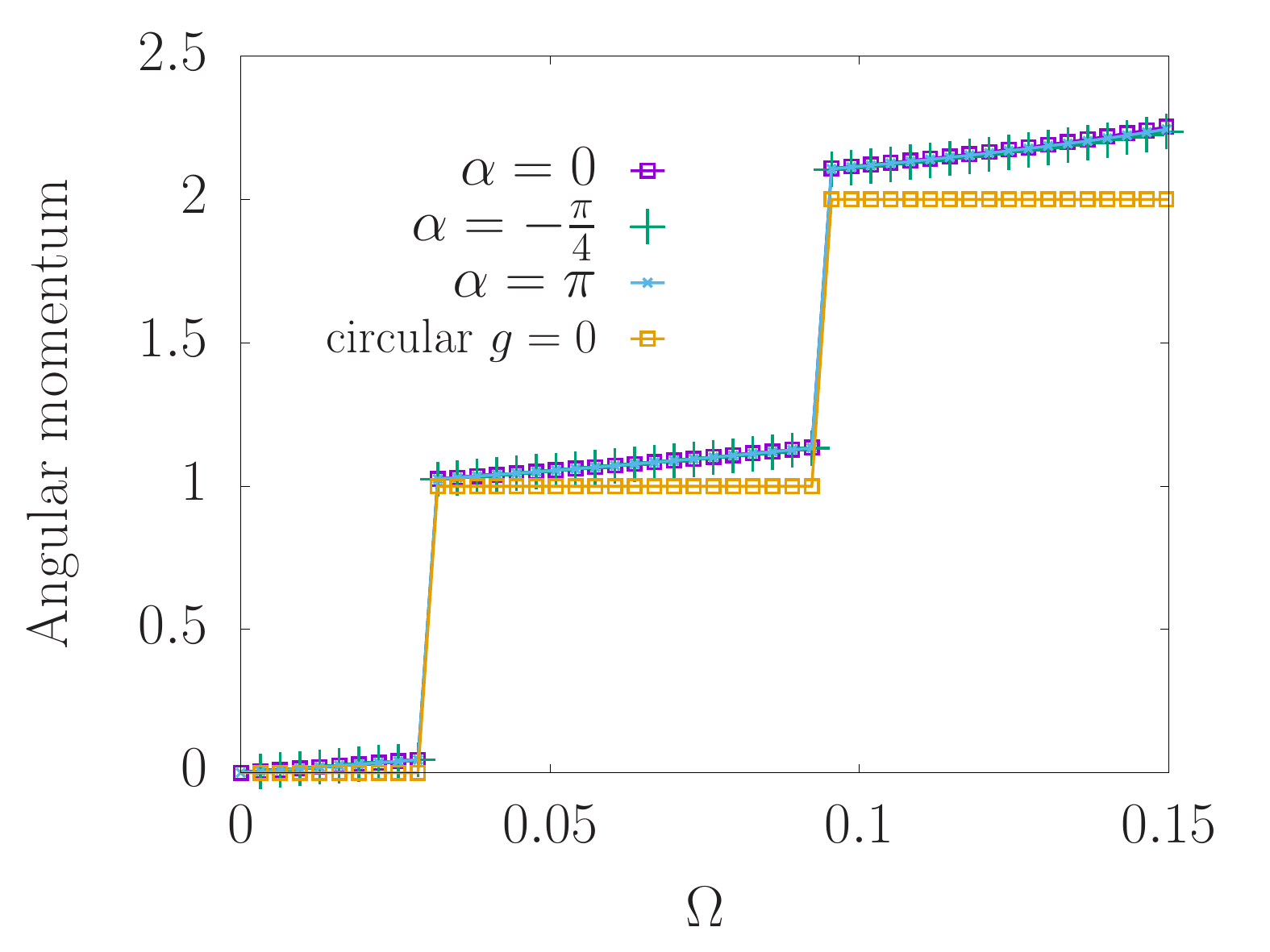}
   \caption{\textbf{Global current} Angular momentum $Ly$ as a function of the induced rotation, $\Omega$, for different values of the angle of the ellipse in a potential as Eq.~\eqref{eq:ellipse}. Parameters of the $2$D GPE are $g=120$, $V_0=15$ and $a/b=\sqrt{1.5}$, $a=25$ and a length scale of $L=30$. For the circular ring we choose a radius of $r\approx5.36$.
   }\label{fig:ring_global_current}
\end{figure}

In this section, we investigate the effects of a global rotation, $\Omega$, on an elliptical ring trap potential. We compare this results with the ones obtained in the main text (see Fig~\ref{fig:Lz_3}). In particular, we consider a GPE of the following form:
\begin{align}
	i \frac{\partial \phi}{\partial t} = \bigg[-\frac{1}{2} \nabla^2 + V_\text{eff} (x,z) + g | \phi |^2 -\Omega L_y \bigg] \phi . \label{GPE_global_rotation}
\end{align}
We note that, in Fig.~\ref{fig:ring_global_current}, all the curves collapse; thus, there is no difference between the various angles of the ellipse. We also add a circular trap for comparison; however, this has a different radius, which produces a small displacement of the critical value that induces a current into the system (see \cite{wright2013driving} for an example of the experimental realization of the plateaus of the current obtained in circular rings).

\section{Continuity equation} \label{sec:Continuity equation}

From the Gross--Pitaevskii equation we can obtain the continuity equation as follows
\begin{equation}
\frac{\partial n}{\partial t} + \nabla \cdot j=0 , \label{eq:J}
\end{equation}
with $j=n(x,z;t) \frac{\hbar}{m} \nabla S(r,t)$, where $n=\psi ^{*} \psi$, is the density of the gas, and $S$ is the phase of the wavefunction, $\psi(r,t) = \sqrt{n(r,t)} e^{i S(x,z;t)}$ \cite{Book_pethick:08}.

For the inhomogeneous case, the equation that governs the system contains a term with an extra spatial dependence, the artificial gauge field (see Eq.~\eqref{GPE}). This leads us to the following equation for $n$ and $j$ with an extra term,
\begin{multline}
 \frac{\partial n}{\partial t} = \frac{i}{2}\Big[\phi^* \nabla^2 \phi -\phi \nabla^2 \phi^* + \\ 
2 i \big( \nabla \cdot \mathbf{A} |\phi|^2 +\phi^* \mathbf{A}.\nabla \phi + \phi \mathbf{A}.\nabla \phi \big) \Big].  \label{New_j}
\end{multline}
In the absence of an inhomogeneous gauge field, by integration over the above equation, we obtain $\oint \nabla S\cdot dl = 2 \pi q$, where $q$ is an integer number. However, the presence of the an inhomogeneous artificial gauge field would prevent us to easily write Eq.~\ref{New_j} in a compact form as $\nabla \cdot \mathbf{A} \neq 0$; For our toy model, 
\begin{equation}
  \nabla \cdot \mathbf{A} = - \frac{w_0}{\beta}\frac{1}
  {(1+z^2/\beta^2)} , 
\end{equation}
this would prevents us from presenting the integral from Eq.~\eqref{eq:J} in a compact form. This equation can be used to explain why, in the elliptical case, the integral depends on the path for the inhomogeneous gauge field. It also indirectly explains how the position-dependent gauge field can affect the value of angular momentum of a vortex added to the cold gas.

\bibliography{bibliography}

\begin{thebibliography}{61}
\bibitem [{HL()}]{HL}%
  \BibitemOpen
  \href@noop {} {}\bibinfo {note} {For a uniform BEC, the size of the core of
  the vortex is the order of the healing length $\xi = (8 \pi na_m)^{-1/2}$,
  where $n$ is the density of the system and $a_m$ is twice of atomic
  scattering length, $a_m = 2a$ \cite{Tempere05}. Also the ratio between $\xi$
  and the radius $r$ of the condensate takes the form of $\xi/R =(a_{ho}/R)^2$
  \cite{Baym96}}\BibitemShut {NoStop}%
\end{thebibliography}

\end{document}